\documentclass[reprint, aps, amssymb, amsmath, prb, superscriptaddress]{revtex4-1}
\usepackage{graphicx}
\usepackage{dcolumn}
\usepackage{bm}
\usepackage{siunitx}
\usepackage{enumerate}
\usepackage[hidelinks]{hyperref}
\usepackage[capitalise]{cleveref}

\usepackage{color}

\begin{document}
\preprint{Draft 7}
\title{Nondegenerate parametric oscillations in a tunable superconducting resonator}
\author{Andreas Bengtsson}
\email{andreas.bengtsson@chalmers.se}
\affiliation{Microtechnology and Nanoscience, Chalmers University of Technology, SE-412 96, G\"oteborg, Sweden}

\author{Philip Krantz}
\affiliation{Research Laboratory of Electronics, Massachusetts Institute of Technology, Cambridge, MA 02139, USA}

\author{Micha\"el Simoen}
\author{Ida-Maria Svensson}
\author{Ben Schneider}
\author{Vitaly Shumeiko}
\author{Per Delsing}
\affiliation{Microtechnology and Nanoscience, Chalmers University of Technology, SE-412 96, G\"oteborg, Sweden}

\author{Jonas Bylander}
\email{jonas.bylander@chalmers.se}
\affiliation{Microtechnology and Nanoscience, Chalmers University of Technology, SE-412 96, G\"oteborg, Sweden}
\date{\today}
\begin{abstract}
We investigate nondegenerate parametric oscillations in a multimode superconducting microwave resonator that is terminated by a SQUID. The parametric effect is achieved by modulating magnetic flux through the SQUID at a frequency close to the sum of two resonator-mode frequencies. For modulation amplitudes exceeding an instability threshold, self-sustained oscillations are observed in both modes. The amplitudes of these oscillations show good quantitative agreement with a theoretical model. The oscillation phases are found to be correlated and exhibit strong fluctuations which broaden the oscillation spectral linewidths. These linewidths are significantly reduced by applying a weak on-resonance tone, which also suppresses the phase fluctuations. When the weak tone is detuned, we observe synchronization of the oscillation frequency with the frequency of the input. For the detuned input, we also observe an emergence of three idlers in the output. This observation is in agreement with theory indicating four-mode amplification and squeezing of a coherent input.
\end{abstract}

\maketitle

\section{Introduction}
The circuit quantum electrodynamics architecture (cQED) \cite{schoelkopf2008wiring} is an attractive platform for quantum information processing with continuous variables.
Within cQED, a variety of nonclassical photonic states can be efficiently generated by  nonlinear superconducting elements --- superpositions of Fock states,\cite{hofheinz2009synthesizing} entangled two-mode photonic states,\cite{eichler2011observation,flurin2012generating,chang2017generating} and multi-photon cat states.\cite{vlastakis2013deterministically}
Parametric phenomena have played important roles in this development.

A typical cQED parametric device consists of a high-quality superconducting resonator integrated with Josephson elements that induce a Kerr nonlinearity in the resonator, and also allow for rapid modulation of the resonator frequency.\cite{sandberg2008tuning,yamamoto2008flux,wallquist2006selective,bergeal2010phase}

By means of such a modulation, at a frequency twice the resonator mode frequency, a degenerate Josephson parametric oscillator (JPO) regime is achieved. \cite{wilson2010photon} The JPO can be used for vacuum squeezing and photonic entanglement, \cite{Wustmann2013,meaney2014quantum} photonic qubit operation, \cite{inagaki2016large} and cat state engineering. \cite{Puri2017,bartolo2016exact,mirrahimi2014dynamically}
The JPO has also been employed for high-fidelity readout of superconducting qubits. \cite{lin2014josephson,philip2016single}

In this paper, we report on an experimental investigation of a different regime, the nondegenerate Josephson parametric oscillator (NJPO). In this regime, self-sustained oscillations of two resonator modes, $n$ and $m$, are excited by modulating the Josephson inductance at a frequency close to the sum of the mode frequencies, $\omega_p \approx \omega_n+\omega_m$.  A detailed theory of nondegenerate parametric resonance has been developed in Ref.~\citenum{Wustmann2017}. The NJPO establishes at parametric pump amplitudes above a critical value, where instability of the resonator ground state develops and is stabilized by the Kerr nonlinearity. Some properties of the nondegenerate parametric resonance in the sub-threshold region --- amplification \cite{simoen2015characterization, chang2017generating} and frequency-conversion \cite{Zakka-Bajjani2011,abdo2013full} have been experimentally investigated.

\begin{figure}[b]
\centering
\includegraphics[width = 1\linewidth]{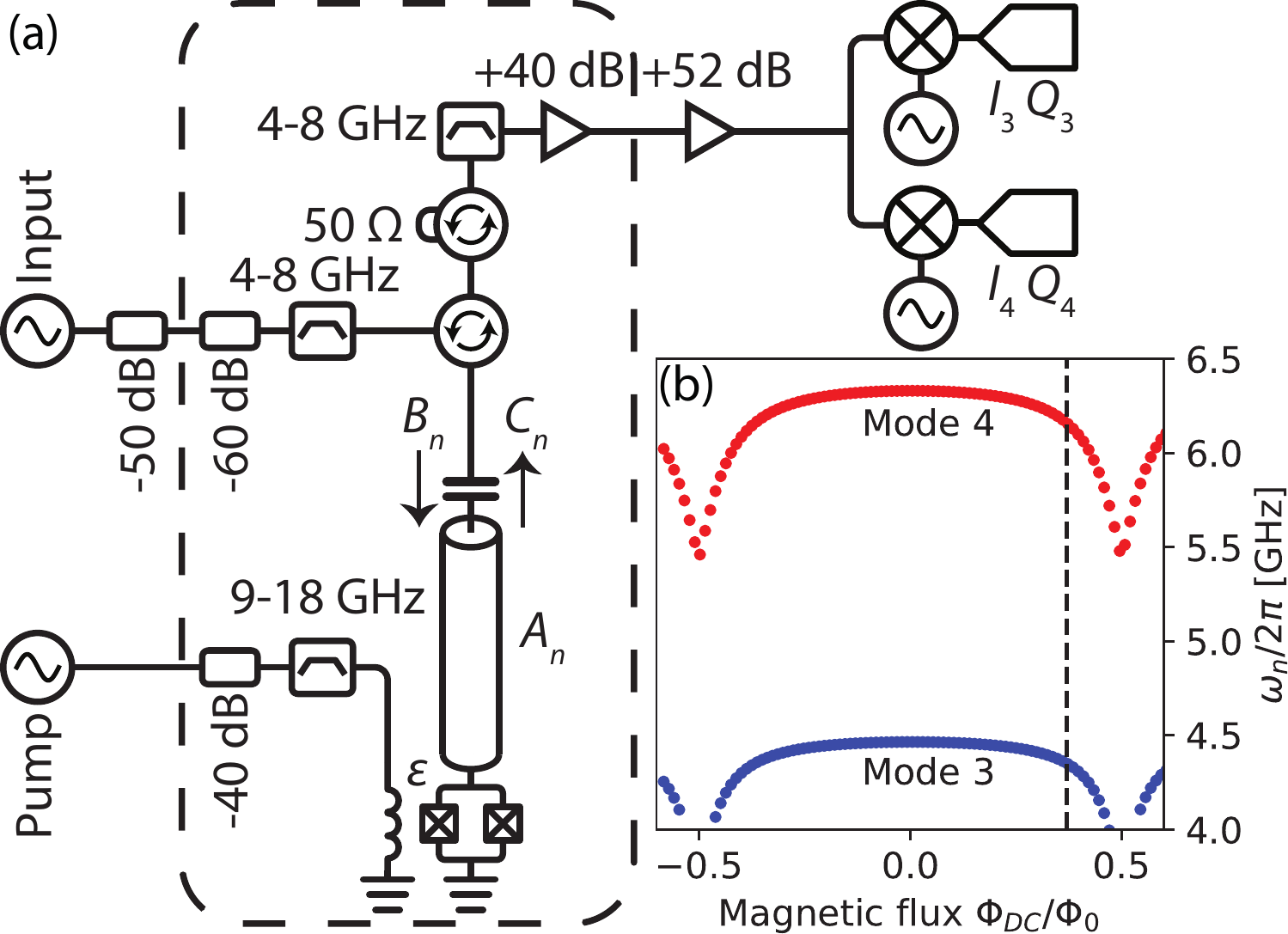}
\caption{\label{fig:setups} (a) Schematic of the experimental setup, where the dashed box indicates the dilution refrigerator. The device consits of a capacitively coupled resonator shorted to ground via a SQUID. The magnetic flux through the SQUID is modulated with an effective amplitude $\epsilon$, via an on-chip microwave line. There are three types of fields present in the system: in-resonator fields $A_n$, input fields $B_n$, and outgoing fields $C_n$, where the last two are separated by a microwave circulator.
The outgoing fields are amplified and split to two vector digitizers measuring the quadratures $I_n$ and $Q_n$.
(b) Measured resonant frequencies for modes 3 (blue) and 4 (red) as a function of magnetic flux through the SQUID loop. The vertical dashed line marks the static flux bias, $\Phi_\mathrm{DC}=0.37\Phi_0$, used to generate nondegenerate parametric oscillations.
}
\end{figure}

Our interest to the NJPO is driven by potentially novel, compared to the JPO, quantum statistical properties of the generated field.
The novelty is related to a presence of additional idlers and multimode squeezing, \cite{Wustmann2017} and large phase fluctuations resulting from a continuous degeneracy of the oscillator state. The latter is analogous to an extensively studied effect in optical parametric oscillators.
\cite{heidmann1987observation,graham1968quantum,mcneil1983quantum,reynaud1987generation,lane1988quantum,fabre1989noise,bjork1988phase,reid1989correlations,courtois1991phase}
To date, neither classical nor quantum properties of the NJPO have been experimentally verified. The aim of this work is to fill this gap. We investigate quasiclassical dynamics of the NJPO: intensity and frequency of the oscillations as functions of the pump parameters, properties of the phase dynamics, and response to external coherent inputs.

\section{Experimental methods}
The device investigated is a coplanar waveguide resonator, capacitively coupled to a transmission line at one end, and shorted to ground via a dc superconducting quantum interference device (SQUID) at the other end, see \cref{fig:setups}(a). The resonator is reactive-ion etched from a sputtered thin-film of niobium on a high-resistivity silicon substrate. The SQUID is deposited with a two-angle evaporation of aluminum.
The layout of the device is similar to that in Ref.~\citenum{simoen2015characterization}. The distance between the interdigitated coupling capacitor and the SQUID is \SI{31}{\milli\meter}, yielding a fundamental resonant frequency $\omega_1/2\pi = \SI{912}{\mega\hertz}$. The available measurement frequency window is 4-8 GHz, limited by the microwave setup, giving experimental access to the higher resonator  modes with numbers 3, 4, and 5. A setup schematic is shown in \cref{fig:setups}(a). The device is mounted at the mixing chamber of a dilution refrigerator with a base temperature of \SI{10}{\milli\kelvin}.

The magnetic field through the SQUID loop can be applied statically via an external coil, or modulated via an on-chip flux line. The resonant frequencies for modes 3 and 4, as functions of the static magnetic flux $\Phi_{\mathrm{DC}}$, are shown in \cref{fig:setups}(b).

Each mode of the NJPO is characterized by the resonant frequency $\omega_n$, external loss rate $\Gamma_{n0}$, total loss rate $\Gamma_{n}$, and Kerr coefficient $\alpha_{n}$. The resonant frequencies and loss rates are determined by data fitting of measured complex reflection coefficients. At the static flux bias ${\Phi_{\mathrm{DC}}=0.37\Phi_0}$ indicated in \cref{fig:setups}(b), we find $\omega_3/2\pi=\SI{4.345}{\giga\hertz}$, $\omega_4/2\pi=\SI{6.150}{\giga\hertz}$, $\Gamma_{30}/2\pi=\SI{0.52}{\mega\hertz}$, $\Gamma_{3}/2\pi=\SI{0.56}{\mega\hertz}$, $\Gamma_{40}/2\pi=\SI{0.70}{\mega\hertz}$, and $\Gamma_{4}/2\pi=\SI{0.78}{\mega\hertz}$.
The Kerr coefficients are determined from properties of the parametric oscillations and explained later.

A microwave signal generator supplies the coherent pump tone to the on-chip flux line via an attenuated and filtered coaxial line. The output of the resonator is routed through microwave filters and two circulators to a cryogenic low-noise amplifier. The output is further amplified at room temperature before detection. Additionally, to investigate the system response, an input field can be applied to resonator via the first circulator.

The quadrature voltages of the output are acquired by heterodyne detection. The use of two sets of digitizers, mixers, and local oscillators, allows for a simultaneous detection of two modes far separated in frequency. From each digitizer, we transfer the digitally downconverted quadrature voltages to a computer for further processing. The output power from each mode, $P_n$, is calculated and related to the output photon flux $|C_n|^2$, ${|C_n|^2  = (P_n - P_{\mathrm{noise}}) / (G \hbar \omega_d)}$, where $P_{\mathrm{noise}}$ is the system noise power measured with the pump off, and $\omega_d$ is the detection frequency. The system gain $G$, is calibrated at each mode frequency using a shot-noise tunnel junction. \cite{spietz2003primary} In a similar way, the quadrature voltages are converted to dimensionless quantities, $I_n(t)$ and $Q_n(t)$, corresponding to the square root of the number of photons per second and unit bandwidth.

\section{Results}
\subsection{Parametric oscillations}
We excite the parametric resonance by modulating the SQUID inductance at a frequency close to the sum of the frequencies of modes 3 and 4, $\omega_p = \omega_3+\omega_4+2\delta$, where $\delta$ refers to the pump detuning. The quantum resonant two-mode dynamics is generally described with a Hamiltonian written in a doubly rotating frame with frequencies  $\omega_{3,4}+\delta$, \cite{Wustmann2013,Wustmann2017}
\begin{align}
\label{eq:H}
 H/\hbar = &- \sum_{n=3,4} \left[ \delta a_n^\dag a_n +  (\alpha_n/2) (a_n^\dag a_n )^2 \right]   \nonumber\\
 & - 2 \alpha (a_3^\dag a_3 a_4^\dag a_4) - \epsilon (a_3 a_4 + a_3^\dag a_4^\dag),
\end{align}
where $a_n$ is the annihilation operator of the in-resonator field of mode $n$, $\epsilon$ is the effective amplitude of the parametric pump, and $\alpha=\sqrt{\alpha_3\alpha_4}$ is the cross-Kerr coefficient.

Throughout this work, we restrict our interpretation of experimental data to a quasiclassical model of the resonator dynamics, which operates with the classical field amplitudes $A_n(t)$. The amplitudes satisfy two dynamical equations,
\begin{align}\label{EOM}
 & i\dot A_3 + (\zeta_3 + i\Gamma_3 ) A_3 + \epsilon A_4^* = \sqrt{2\Gamma_{30}} B_3(t),
\nonumber\\
& i\dot A_4 + (\zeta_4 + i\Gamma_4  ) A_4 + \epsilon A_3^* = \sqrt{2\Gamma_{40}} B_4(t),
\end{align}
where $B_n(t)$ is an external driving field, and $\zeta_n$ is a nonlinear detuning including the Kerr-induced frequency shifts,
\begin{align}\label{eq:zeta}
\zeta_3 &= \delta + \alpha_3|A_3|^2 + 2\alpha|A_4|^2, \nonumber\\
\zeta_4 &= \delta + \alpha_4|A_4|^2 + 2\alpha|A_3|^2.
\end{align}
The normalization of the field amplitudes, $|A_n|^2$ and $|B_n|^2$, corresponds to the number of photons in respective mode and the incoming photon flux, respectively. Parametric self-sustained oscillations correspond to nontrivial solutions of the homogeneous nonlinear equations \eqref{EOM} and \eqref{eq:zeta}.

\begin{figure}
\centering
\includegraphics[width = 1\linewidth]{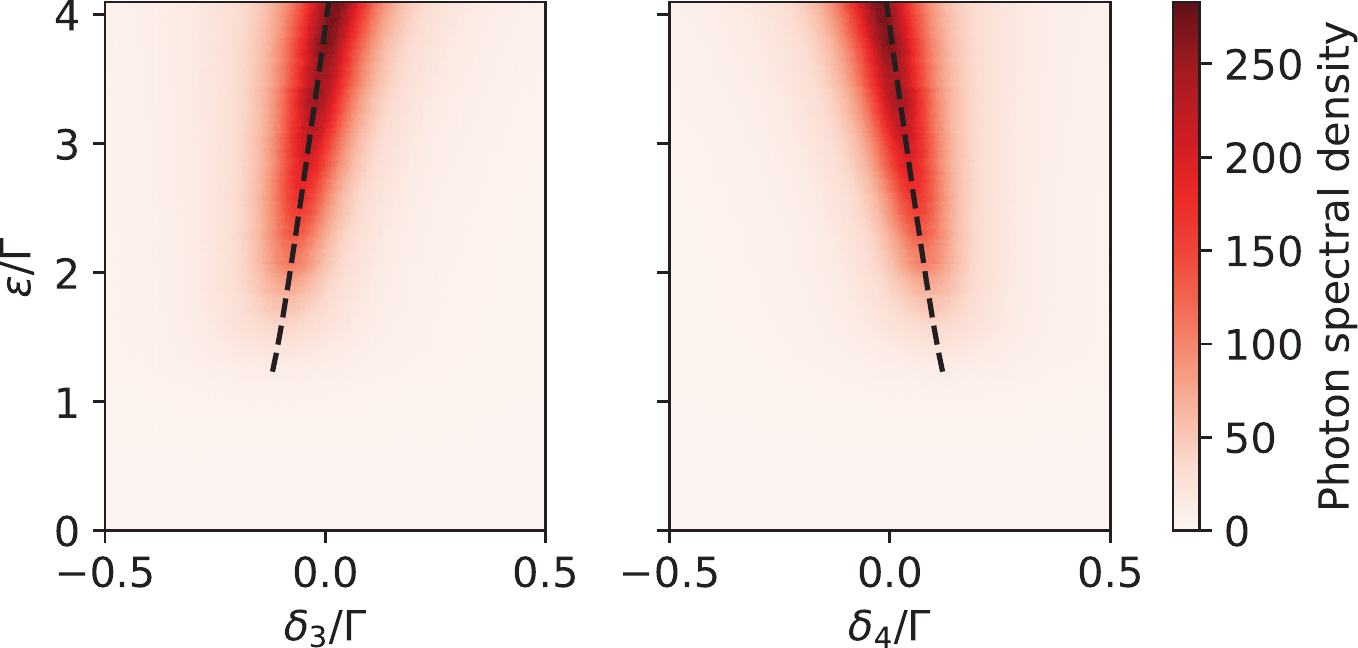}
\caption{\label{fig:ndpo}
Parametric instability. Measured photon spectral densities of the output from mode 3 (left) and 4 (right). The parametric pump has an amplitude $\epsilon$, and is applied at a detuning $\delta=0.26\Gamma$. The detection frequency detunings $\delta_n$, is relative to the respective rotating frame. The dashed lines are the radiation frequencies from \cref{eq:Delta_0}.}
\end{figure}

\subsubsection{Parametric instability}
By ramping the pump amplitude we observe a strong increase of the output photon flux above a certain pump threshold, as shown in \cref{fig:ndpo}. The radiation is detected at frequencies $\omega_d$, associated with the excited resonator modes but deviating from the respective rotating frames,  $\delta_n = \omega_d-(\omega_n+\delta)$, in good agreement with a theoretical prediction, \cite{Wustmann2017}
\begin{align}\label{eq:delta_n}
\delta_3 = -\delta_4 = \delta \frac{\Gamma_3- \Gamma_4}{\Gamma_3+\Gamma_4}.
\end{align}
With further increase of the pump power, the radiation frequencies shift, as shown in \cref{fig:ndpo}. This shift is accurately described by the equation, \cite{Wustmann2017}
\begin{align}\label{eq:Delta_0}
\delta_3(\epsilon) = -\delta_4(\epsilon) = \Delta_0 = \frac{\Gamma_3\zeta_4-\Gamma_4\zeta_3}{\Gamma_3+\Gamma_4}.
\end{align}

\begin{figure}
\centering
\includegraphics[width = 1\linewidth]{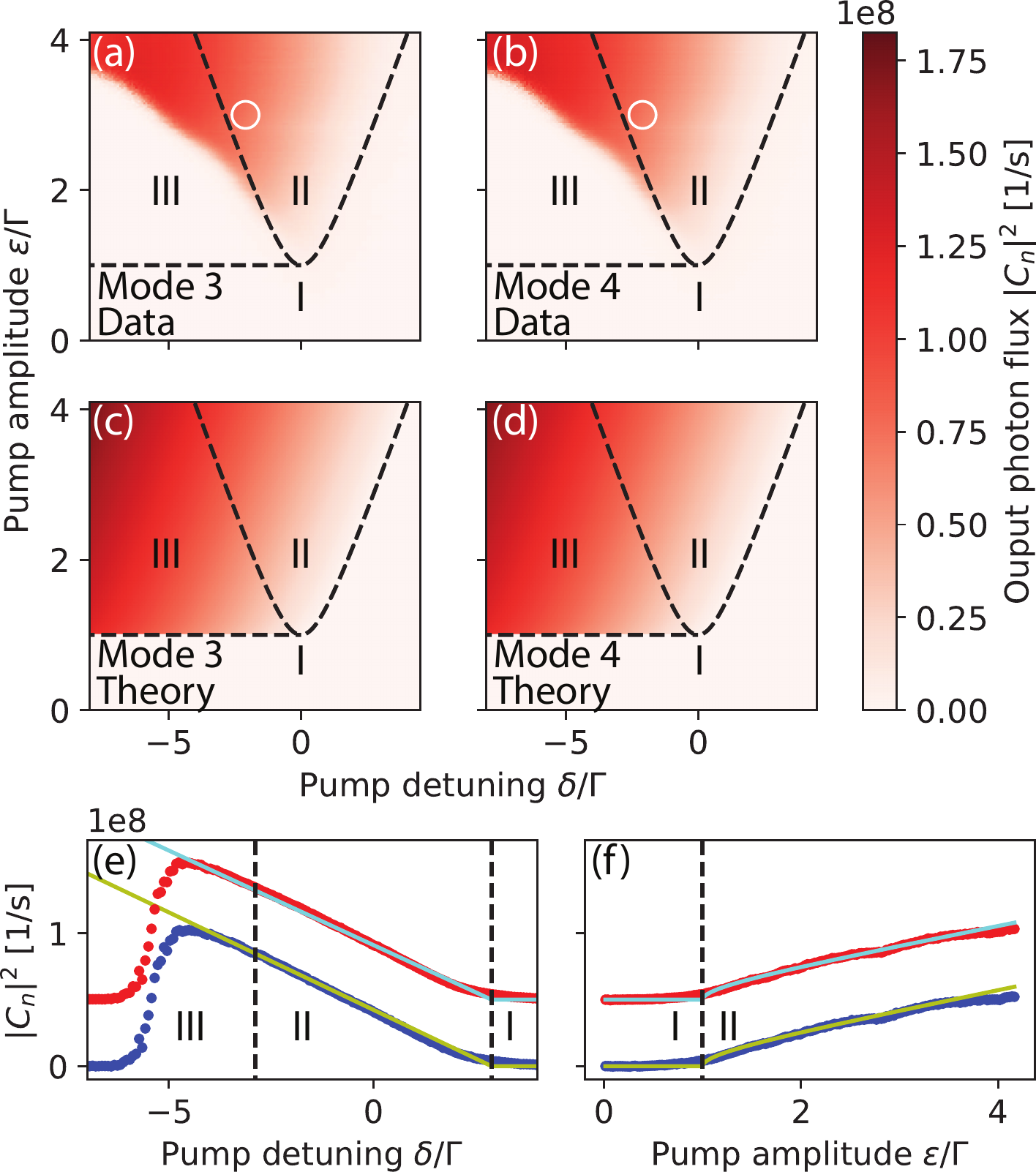}
\caption{\label{fig:ndpo_amp}Nondegenerate pumping of modes $n=3$ and 4, at the frequency $\omega_p=\omega_3+\omega_4+2\delta$.  Experimentally measured (a-b), and theoretically calculated (c-d), output intensities $|C_{n}|^2$ vs.~pump detuning $\delta$ and amplitude $\epsilon$. I-III indicate the three different stability regions described in the main text. (e) Horizontal line cuts of (a-d) at $\epsilon=3\Gamma$, where mode 4 is offset in the positive y-direction for clarity. The dots are measured values and solid lines are from theory. (f) Vertical line cuts of (a-d) at $\delta=0$, where mode 4 is offset in the positive y-direction for clarity. The dots are measures values and solid lines are from theory.}
\end{figure}

The instability of the resonator ground state occurs within an interval of the pump detuning,
\begin{equation}
\label{eq:P}
|\delta| < \delta_{\mathrm{th}}(\epsilon) = \frac{\Gamma_3+\Gamma_4}{2}\sqrt{\frac{\epsilon^2}{\Gamma^2}-1}.
\end{equation}
This criterion defines three regions in the $\epsilon$-$\delta$ plane, as presented in \cref{fig:ndpo_amp}:\\
(I) at $\epsilon<\Gamma$ or $\delta > \delta_{\mathrm{th}}(\epsilon)$ only the ground state, $A_n=0$, is stable;\\
(II) at $\epsilon>\Gamma$ and $|\delta| < \delta_{\mathrm{th}}(\epsilon)$ the ground state is unstable and self-sustained oscillations emerge;\\
(III) at $\epsilon>\Gamma$ and $\delta < -\delta_{\mathrm{th}}(\epsilon)$ the ground state regains stability while the self-sustained oscillations persist --- this is a bistability region.

\subsubsection{Output intensities}
A quantitative analysis of the intensity of the oscillations is performed by solving \cref{EOM},
\begin{align}
\label{eq:A}
|A_{3}|^2 &= \frac{2\Gamma_4(\delta_{\mathrm{th}}(\epsilon)-\delta)}{\alpha_3\Gamma_4+\alpha_4\Gamma_3+2\alpha(\Gamma_3+\Gamma_4)} , \\
 |A_{4}|^2 &= \frac{\Gamma_3}{\Gamma_4}|A_{3}|^2 .
\end{align}
The output intensity is given by the relation ${|C_n|^2 = 2\Gamma_{n0}|A_n|^2}$; for the experimentally extracted external and total losses, ${|C_3|^2 \approx |C_4|^2}$.

In \cref{fig:ndpo_amp}, the measured output intensities are shown on panels (a) and (b), while panels (c) and (d) show the computed intensities of the oscillations. In the bistability region III, the output intensities decrease with red detuning, indicating a system preference to occupy the ground state.
Measured output intensities, as a function of the pump detuning at a fixed pump amplitude $\epsilon=3\Gamma$, are presented in \cref{fig:ndpo_amp}(e). The data and the theory are in excellent agreement in region II and down to $\delta\approx -4.5\Gamma$ in region III, implying that the system is mostly in the excited state.
Below $\delta\approx -4.5\Gamma$, the occupancy of the excited state rapidly decreases. \Cref{fig:ndpo_amp}(f) illustrates the growth of the output intensities with the pump amplitude at $\delta=0$, also showing good agreement with the theory.

The slopes of the frequency shift in \cref{fig:ndpo} and the output intensity in \cref{fig:ndpo_amp}(e), together with \cref{eq:Delta_0,eq:A}, are used to extract the Kerr coefficients as well as the conversion coefficient between the flux pump amplitude and $\epsilon$.
The extracted values for the Kerr coefficients are $\alpha_3/2\pi=\SI{71}{\kilo\hertz}$ and $\alpha_4/2\pi=\SI{178}{\kilo\hertz}$.

\subsubsection{Phase dynamics}
\begin{figure}
\includegraphics[width = 1\linewidth]{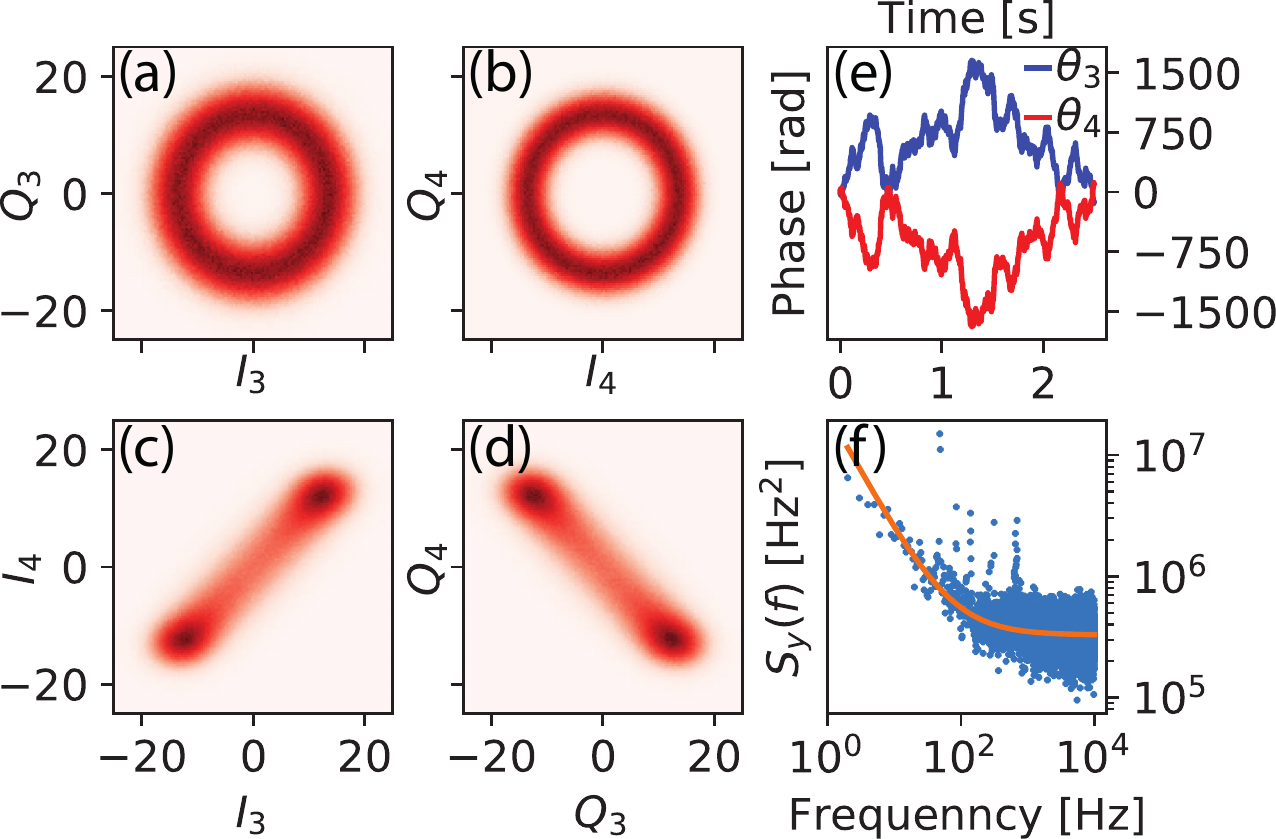}
\caption{\label{fig:ndpo_hist}Phase-space distributions for nondegenerate parametric oscillations measured at the point indicated by the white circles in \cref{fig:ndpo}. Panels (a-b) show the distributions for modes 3 and 4. The color scale is proportional to the number of counts in each bin. (c-d) Two out of four cross-quadrature histograms, showing clear anti-correlation between the mode phases $\theta_n$.
(e) Evolution of the phases $\theta_n$ in time. (f) Frequency-noise spectrum of the parametric oscillations in mode 3. The solid line shows a combination of $1/f$ and white noise.
}
\end{figure}

We further investigate the phase properties of the parametric oscillations. To this end, we choose the point in the $\delta$-$\epsilon$ space indicated by the white circles in \cref{fig:ndpo_amp}(a-b), and acquire 1 million samples for the quadratures $I_n(t)$ and $Q_n(t)$, during 2.5 seconds of measurement time. By creating two-dimensional histograms, we present the data in \cref{fig:ndpo_hist}(a-b). The oscillations have a finite average amplitude, while the phase is evenly distributed between $-\pi$ and $\pi$. This observation supports the theoretical prediction about a continuous degeneracy of the oscillator state with respect to the phase.\cite{Wustmann2017} More precisely, the oscillator phases
$|A_n|e^{i\theta_n}$, respect the constraint,
\begin{align}\label{eq:phases}
&\theta_3 + \theta_4 = \Theta, \quad \Theta \in \{\pi/2, \pi \}, \nonumber \\
& \tan\Theta = -\frac{1} {\sqrt{\epsilon^2/\Gamma^2-1}},
\end{align}
while the difference of the phases, $\psi = \theta_3 - \theta_4$, is arbitrary. Such a degeneracy gives rise to phase diffusion under the effect of vacuum fluctuations,
which underlines the broadening of the spectrum of the output signal in \cref{fig:ndpo}.

To reveal the intermode phase correlation, we synchronize the digitizers by using a common trigger. This allows us to create the cross-quadrature histograms, {$I_3,\, I_4$} and ${Q_3,\, Q_4}$, in an analogous way to the phase-space distributions. \cref{fig:ndpo_hist}(c-d) present the cross-quadrature histograms for quadratures chosen such that their common output phase $\Theta$ is compensated by the phases of the local oscillators. With such a choice, the histograms exhibit the relations, $I_3=I_4$, and $Q_3= -Q_4$.
To further illustrate the phase anti-correlation property, we plot in \cref{fig:ndpo_hist}(e), a time evolution realization of the phases of the two modes.

The effective frequency-noise spectrum can be extracted from the phase evolution. It is presented for mode 3 in \cref{fig:ndpo_hist}(f). The spectrum is in good agreement with a $1/f$ component combined with white noise. The origin of the low-frequency noise is most likely due to flux noise through the SQUID loop, which is known to have a $1/f$ spectrum. \cite{koch1983flicker}

\subsection{Response to an external signal}
In this section, we explore the response of the NJPO to an external coherent input. Linear response of parametric systems to weak external signals underlines many properties of quantum noise. For instance, the presence of an idler in parametric amplification below the threshold, defines the structure of squeezed vacuum and two-mode entanglement of output photons.\cite{eichler2011observation,flurin2012generating,chang2017generating}  This is true for both degenerate and nondegenerate parametric resonance, with the only difference that for the nondegenerate case, the idler has a frequency far detuned from the signal frequency and appears within the bandwidth of the conjugated mode,\cite{simoen2015characterization} while for the degenerate case, the idler appears within the bandwidth of the signal mode.

\begin{figure}
\includegraphics[width = 1\linewidth]{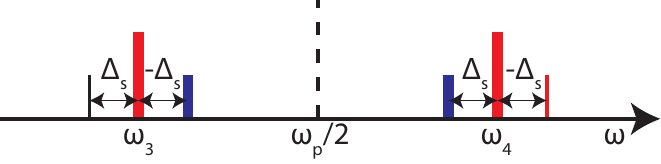}
\caption{\label{fig:idlers}
Frequency diagram of the four-mode amplification. The dashed line is the parametric pump, the bold red lines are the parametric oscillations, the thin black line is the coherent signal, the thin red line is the primary idler, and the bold blue lines are the secondary idlers.}
\end{figure}

Above the threshold, the situation is qualitatively similar for the JPO.\cite{Wustmann2013} However, for the NJPO, the situation is quite different. Here, the strong fields produced by the parametric oscillations in the two resonator modes generate, through a four-mode mixing mechanism, two additional idlers, \cite{Wustmann2017} see \cref{fig:idlers}.
The intensity of these secondary idlers are proportional to the oscillation intensities $|A_n|^2$, while the intensity of the primary idler is defined by, and proportional to, the flux pump intensity $\epsilon^2$. This process of four-mode amplification should result in a four-mode quantum noise squeezing. The output noise should also be influenced by the strong fluctuations of the oscillation phases discussed above.

In this section, we present data that corroborate the presence of the three idlers in the NJPO response. In addition, we observe two effects that imply a strong influence of the input signal on the oscillator phase dynamics --- phase locking, and frequency synchronization.

\subsubsection{Injection locking}
It is generally known that in self-sustained oscillators possessing phase degeneracy,
large phase fluctuations can be suppressed by injecting a small, but frequency stable, signal on-resonance with the oscillator.\cite{adler1946study}
This effect is explained by a violation of the symmetry underlining the phase degeneracy by external driving. The phase locking effect has been observed in a nondegenerate optical parametric oscillator. \cite{jing2006experimental} For our system, it has been shown,\cite{Wustmann2017} that applying an input with the same frequency as the oscillator frequency, locks the oscillator phase to the value defined by the phase of the input, $\theta_{\mathrm{in}}$,
\begin{equation} \label{eq_psi}
\theta_3 =   \theta_{\mathrm{in}}  - \arctan \frac{3\Gamma}{2\sqrt{\epsilon^2-\Gamma^2}}.
\end{equation}
Due to the intermode phase coupling, \cref{eq:phases}, both modes are phase locked simultaneously.

\begin{figure}
\includegraphics[width = 1\linewidth]{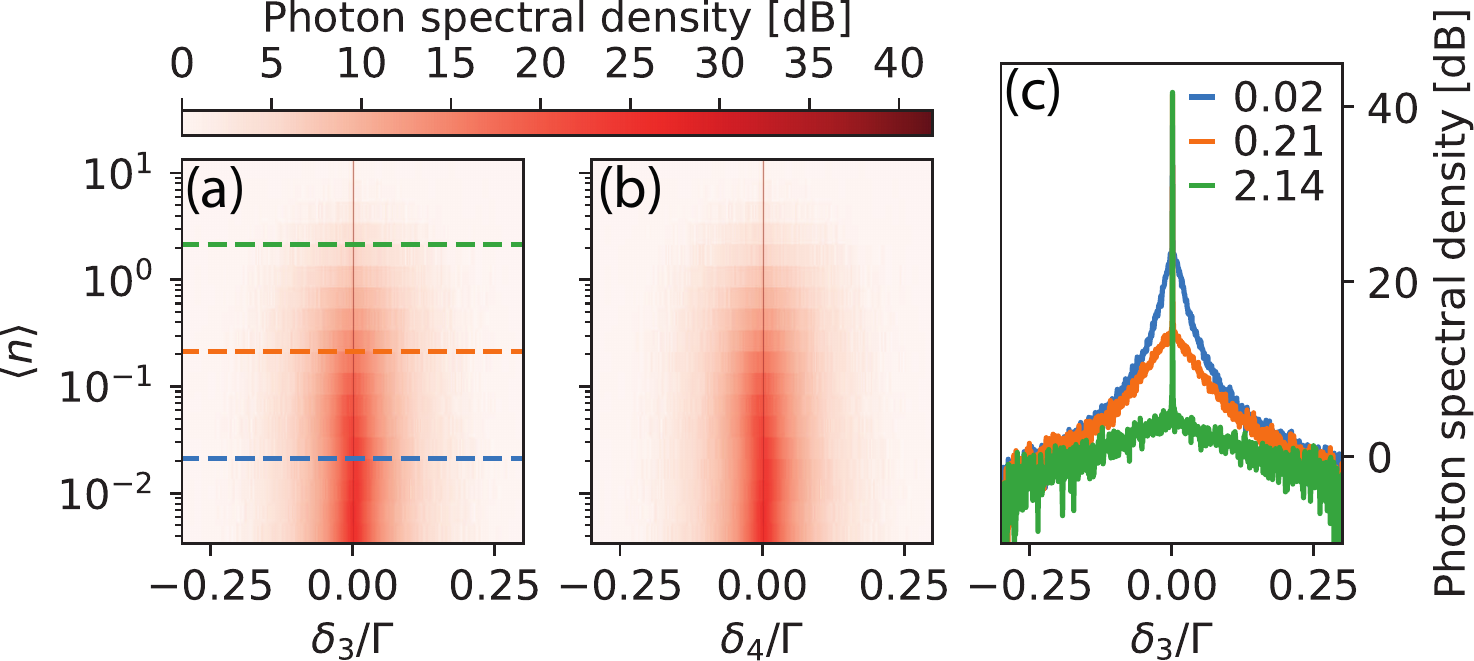}
\caption{\label{fig:ndpo_locking}Photon spectral densities of a nondegenerate parametric oscillator under injection locking. The scale is presented in a logarithmic unit relative to 1 photon/(sHz). Panels (a) and (b) show the output radiation for modes 3 and 4, respectively. $\delta_n$ is the detuning between the detection frequency and the center frequency of the radiation in the respective mode, and $\langle n \rangle$ is the average input photon number to mode 3. (c) Line cuts of (a) at the three input photon numbers indicated by the corresponding colored dashed lines in (a).}
\end{figure}

We investigate this locking effect by injecting a coherent signal $B_3$ on resonance with the parametric oscillations in mode 3. We characterize the input field with an average number of coherent photons $\langle n \rangle = |B_3|^2/(2\Gamma_{30})$.
To quantify the efficiency of the injection locking, we measure the output spectral densities as functions of the input photon number $\langle n \rangle$, which is presented in \cref{fig:ndpo_locking}(a-b). In \cref{fig:ndpo_locking}(c), we plot line cuts for mode 3 for several injection photon numbers.
For low input photon numbers, the radiation line is still broad, but when the input power is increased, the width is substantially reduced, and if increased further, the frequency noise is removed almost entirely. The \SI{-3}{\decibel} point is below the resolution bandwidth of \SI{1}{\hertz}, implying a frequency noise reduction of at least a factor 5000. This narrowing effect becomes pronounced at $\langle n \rangle \geq 0.5$, similarly to what was found to phase lock the JPO. \cite{lin2014josephson}
This result can be understood from the following argument: under the effect of vacuum noise, the oscillator phase undergoes random motion, which is shown in \cref{fig:ndpo_hist}(f) as the white frequency noise above \SI{100}{\hertz}. This random motion can only been constrained by a coherent input whose strength exceeds the intensity of the vacuum fluctuations, $\langle n\rangle=1/2$.

\begin{figure}
\includegraphics[width = 1\linewidth]{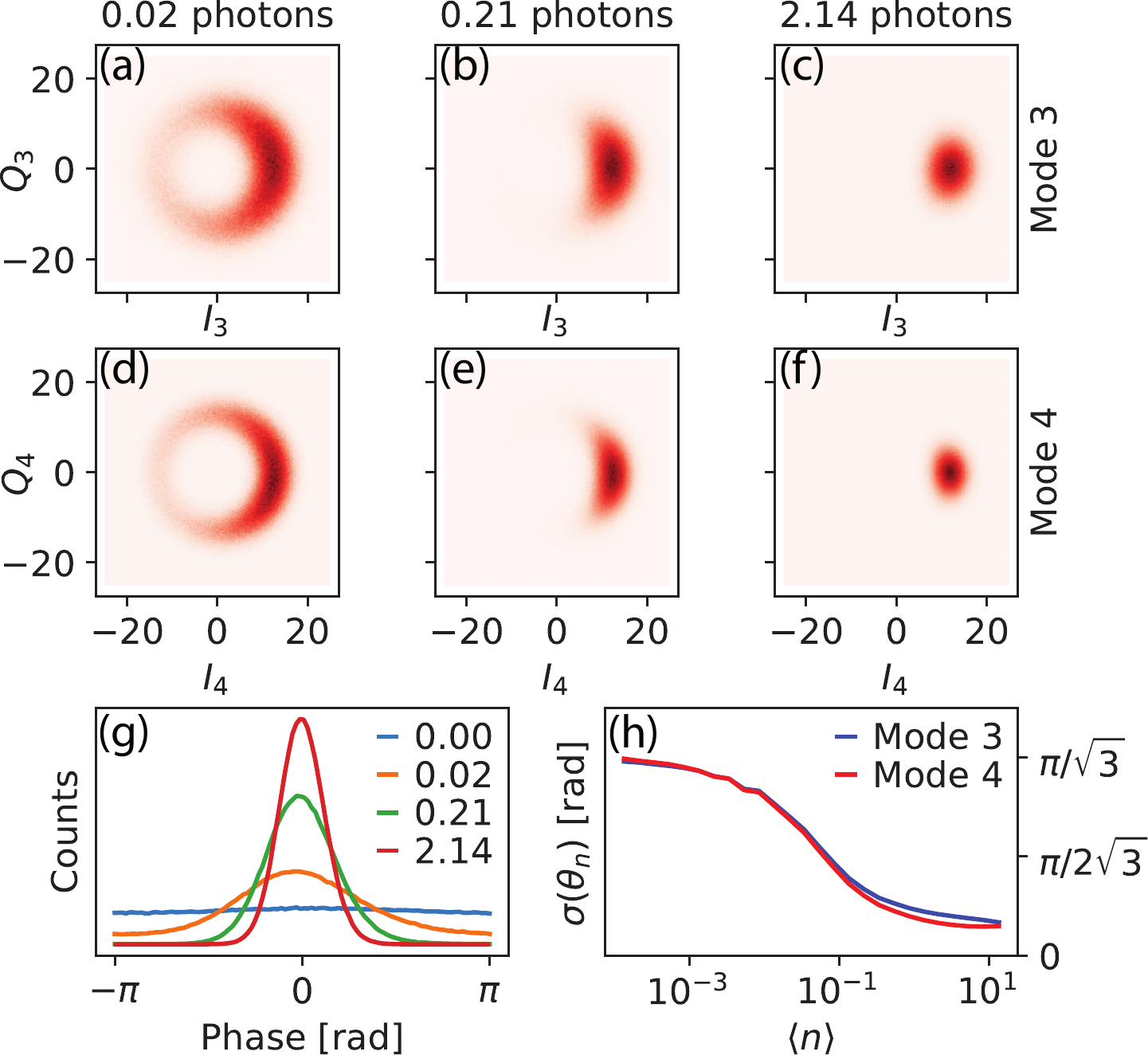}
\caption{\label{fig:ndpo_tilt}Phase locking of nondegenerate parametric oscillations by signal injection. Phase-space histograms for the parametric radiation in modes 3 (a-c) and 4 (d-f), with a coherent signal injected into mode 3. The input photon number $\langle n \rangle$ is stated above each panel column, and are the same as in \cref{fig:ndpo_locking}(c). The color scale is proportional to the number of counts in each bin. (g) Distributions of the phase of the parametric oscillations in mode 3, $\theta_3$, for different input photon numbers. (h) Standard deviation of the phases $\theta_n$, as functions of the input photon number to mode 3.}
\end{figure}

\Cref{fig:ndpo_tilt}(a-f) illustrates the phase-space distributions for both modes at the same input powers as in \cref{fig:ndpo_locking}(c). The calculated phase distributions and standard deviations are presented in \cref{fig:ndpo_tilt}(g) and (h), respectively, as functions of the input photon number. The phase distribution is uniform between $-\pi$ and $\pi$ for small injection signals, and has a standard deviation close to that expected for a uniform distribution, $\pi/\sqrt{3}$. For increasing $\langle n \rangle$, the distribution approaches a Gaussian form, and the standard deviation saturates. It is difficult to compare the eventually locked phase with the theoretical value, \cref{eq_psi}, because of the difficulty of experimentally calibrating the phase accumulation between the resonator and the detectors.

\subsubsection{Synchronization and secondary idlers}
\begin{figure}
\includegraphics[width = 1\linewidth]{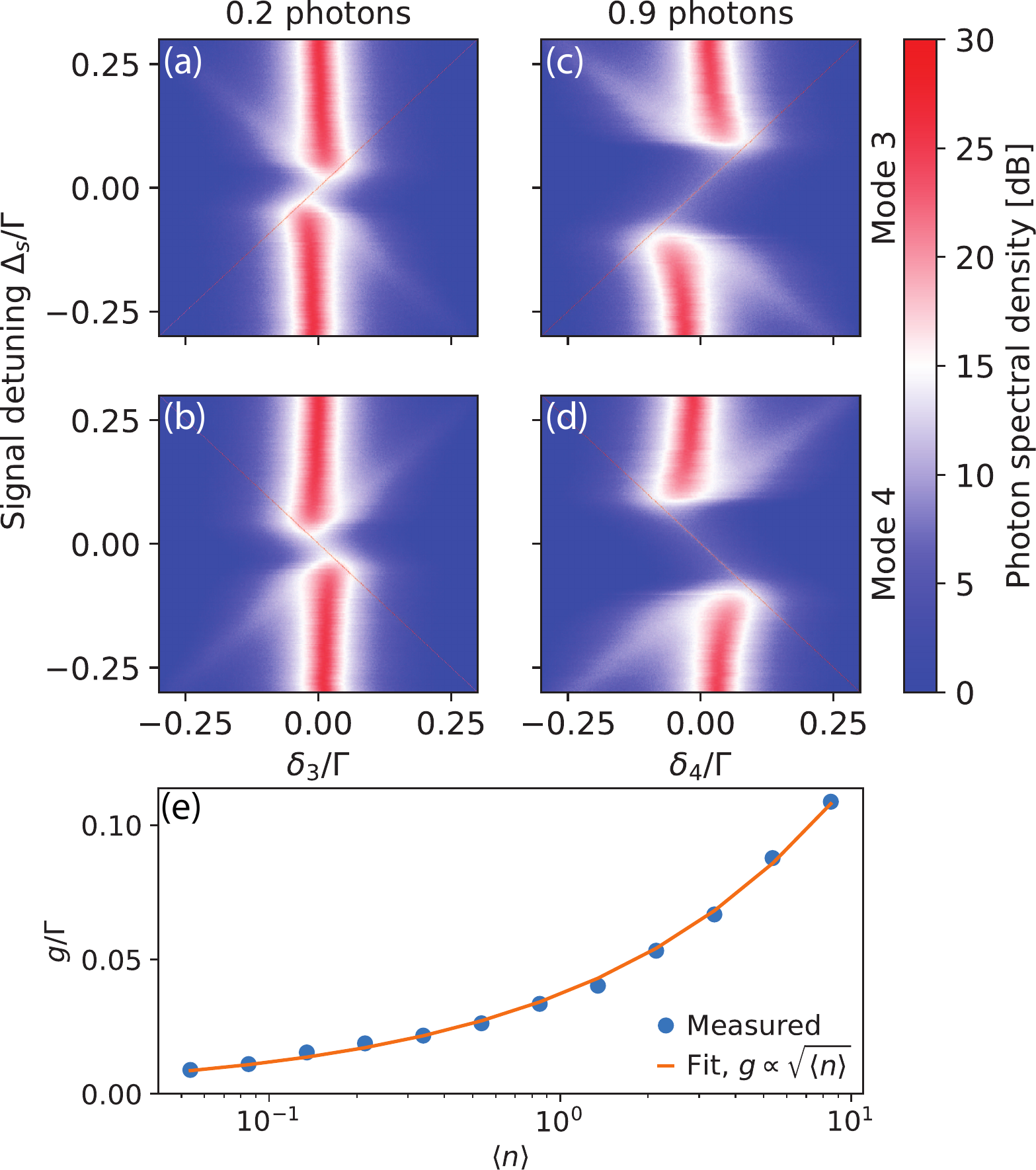}
\caption{\label{fig:ndpo_idlers}
Response of the parametric oscillations to a detuned signal. Panels (a-d) show the output photon spectral densities on a logarithmic scale, for mode 3 (a,c) and mode 4 (b,d). The detection detunings $\delta_n$ are relative to from the center of the parametric oscillations in the respective mode. The signal detuning $\Delta_s$, is the detuning between the coherent signal and the center of the oscillation frequency in mode 3. The input photon number is stated above the two panel columns. Panel (e) shows the obtained size of the frequency gap $g$, and a fit to $g\propto\sqrt{\langle n \rangle}$.}
\end{figure}
Applying an input signal detuned from the oscillation frequency, gives rise to the interesting and related phenomenon of frequency synchronization. \cite{adler1946study,pikovsky2003synchronization}
We introduce a detuning $\Delta_s$ between the injection signal and the parametric oscillations in mode 3, and study the output photon spectral densities of both modes as functions of $\Delta_s$ and $\langle n \rangle$, see \cref{fig:ndpo_idlers}(a-d). Within a certain interval of detuning, we observe a sudden change of the oscillation frequency, which synchronizes with the frequency of the input. Simultaneously, the frequency of the conjugated mode synchronizes with the frequency of the primary idler. Due to the dramatic decrease of the linewidth of the synchronized output signal, the effect appears in \cref{fig:ndpo_idlers}(a-d) as a gap in frequency space where the oscillations have the same frequency as the locking signal. The gap size is proportional to the size of the synchronization detuning window, which in turn is proportional to the square root of the input power. \cite{adler1946study} In \cref{fig:ndpo_idlers}(e), we quantify the gap size and find good agreement with the predicted square-root dependence.

\Cref{fig:ndpo_idlers}(a-d) also reveals the presence of three idlers in the output. The output signal in mode 3, seen as a thin diagonal line on panels (a) and (c)
(hardly visible at the small intensity on panel (a)) generates a primary idler in mode 4, seen as a thin diagonal line in the opposite direction compared to the signal, on panels (b) and (d). This idler has its frequency detuned by $-\Delta_s$ from the oscillation frequency in this mode, and it has a small linewidth, similar to the signal. The two secondary idlers are visible on panels (a) to (d). The secondary idler on panels (a) and (c) is detuned by $-\Delta_s$ from the oscillation frequency of mode 3, while the secondary idler on panels (b) and (d) is detuned by $\Delta_s$ from the oscillation frequency of mode 4. These secondary idlers are generated by the in-resonator fields of the parametric oscillator, and they are much broader than the signal and primary idler, their linewidths are instead comparable to the one of the oscillations.

\section{Conclusion}
In conclusion, we investigated a nondegenerate Josephson parametric oscillator, NJPO, using a tunable superconducting resonator.
By modulating magnetic flux through the SQUID attached to the resonator, we generated intense correlated output radiation of two resonator modes. The measured radiation frequencies and intensities, as functions of the pump parameters, show excellent quantitative agreement with theory. A correlated phase dynamics of the oscillations was directly observed, and a continuous phase degeneracy of the oscillations was demonstrated. We also demonstrated significant suppression of the phase fluctuations when a weak on-resonance coherent signal was applied.
Simultaneously, the oscillation linewidths were reduced by at least three orders of magnitude. In addition, the frequency synchronization effect was observed when the input signal was detuned from the resonance. Such an input was found to generate three output idlers in agreement with theoretical predictions.

Our findings form a solid ground for further exploration of the quantum properties of the NJPO field, that would exhibit four-mode squeezing and might possess non-Gaussian properties.

\begin{acknowledgments}
We wish to express our gratitude to Waltraut Wustmann, Giulia Ferrini, G\"oran Johansson, and Jonathan Burnett for helpful discussions. We also thank Jos\'e Aumentado for providing the shot-noise tunnel junction. We acknowledge financial support from the Knut and Alice Wallenberg foundation, and from the Swedish Research Council. J.B.~acknowledges partial support by the EU under REA grant agreement no.~CIG-618353.
\end{acknowledgments}

\bibliography{njpo_v1.bbl}
\end{document}